\documentclass[doublespacing]{elsart}
\usepackage{amsmath,amssymb}
\usepackage{graphicx}
\usepackage{subfigure}

\begin{document}
\begin{frontmatter}
\title{The boiling suppression of liquid nitrogen}

\author{S. Takayoshi\corauthref{cor}},
\corauth[cor]{Corresponding author. Tel.: +81 3 5841 8364; fax: +81 3 5841 8364.}
\ead{takayoshi@kelvin.phys.s.u-tokyo.ac.jp}
\author{W. Kokuyama},
\author{H. Fukuyama}

\address{Department of Physics, Graduate School of Science, The University of Tokyo, 
7-3-1 Hongo, Bunkyo-ku, Tokyo 113-0033, Japan}

\begin{abstract}
When He gas is injected from room temperature into boiling liquid N$_{2}$,
boiling is suppressed, leaving liquid surface flat like a mirror.
Although the qualitative explanation for this phenomenon is known
[Minkoff G J \textit{et al}.
Nature 1957;180(4599):1413-4.],
it has not been studied quantitatively and comprehensively yet. 
In this report, we made careful simultaneous measurements of
temperature and weight variation of the liquid.
The results clearly indicate that the boiling suppression is caused by cooling of the
liquid with ``internal evaporation'' of N$_{2}$ into the He bubbles.
\end{abstract}
\end{frontmatter}

\section{Introduction}
\begin{figure}[thbp]
\begin{center}
\includegraphics[width=100mm]{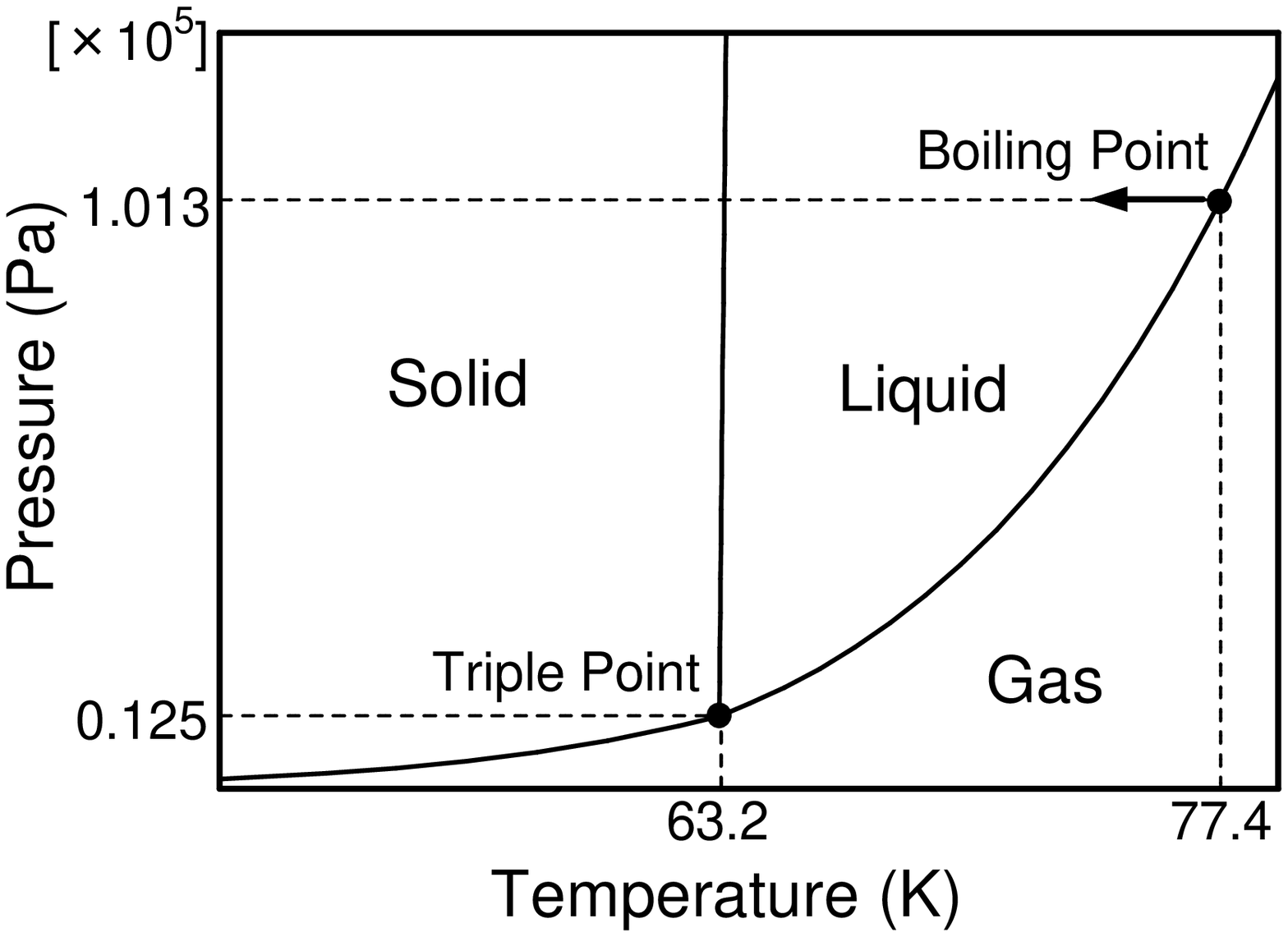}
\end{center}
\caption{The phase diagram of N$_{2}$. The liquid is cooled along the arrow.}
\label{fig:1}
\end{figure}

Boiling liquid N$_{2}$ is a common and useful cryogen. However, the boiling does
various harms to experiments. For example, it makes the liquid a non-uniform medium
for optical experiments and generates vibrational disturbances to the system.
The boiling can be stopped by injecting He gas from room temperature
into the liquid. It was reported that
the cause of this phenomenon is the ``internal evaporation'',
i.e., liquid N$_{2}$ is forced to evaporate into injected He bubbles
until the evaporation pressure reaches the equilibrium \cite{Minkoff, Xu}.
Because the latent heat is deprived by the evaporation,
the liquid is cooled down.
He bubbles mixed with N$_{2}$ gas rise up to the air,
and new bubbles are supplied one after another,
lowering the liquid temperature.
Boiling is suppressed by this temperature dropdown,
not by the pressure dropdown,
as shown by the arrow in the Fig. \ref{fig:1}.
However, this scenario has not been examined quantitatively enough.
We studied this phenomenon quantitatively and comprehensively,
which confirmed the above-mentioned scenario unambiguously \cite{Takayoshi}.

\section{Methods}
\begin{figure}[thbp]
\begin{center}
\includegraphics[width=100mm]{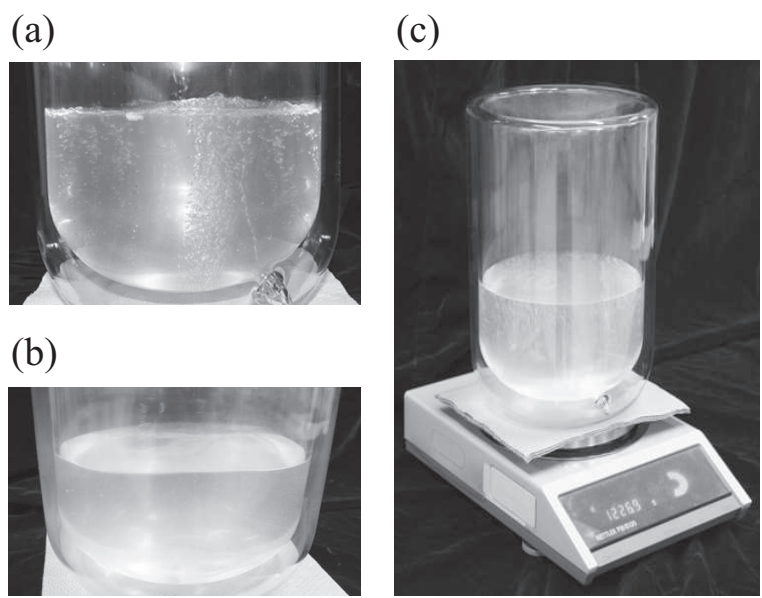}
\end{center}
\caption{(a)Boiling liquid N$_{2}$. Bubbles are formed on the inner
wall of the glass dewar.
(b)After the injection of He gas. Boiling is
suppressed and the surface is quiet like a mirror.
(c)The experimental setup of the weight measurement of liquid N$_{2}$
by an electronic scale.}
\label{fig:2}
\end{figure}

Let us describe our experimental setup and procedure. We used a double-walled glass dewar
(inner diameter 150 mm, depth 240 mm) without silver-plating. When liquid N$_{2}$ is poured
into the dewar, violent boiling happens first. After a while, the inner wall of the
dewar is cooled down and liquid N$_{2}$ boils steadily as shown in the photograph (Fig. \ref{fig:2}(a)).
The bubbles are formed one after another at nucleation centers, such as
scratches and dust on the inner wall surface.
Then He gas was injected into the liquid N$_{2}$ through a
stainless steel pipe
(5 mm in diameter), whose end was placed near the bottom of the dewar. 
The flow rate of injected gas was measured by a flow meter with a 5 \% precision.
After the injection, the boiling is suppressed and the liquid surface becomes flat
like a mirror as seen in the photograph (Fig. \ref{fig:2}(b)).
The suppression continues for a while and finally the boiling starts again.
This phenomenon is reproducible, and the suppression time depends on the amount of injected He gas.

We used a carbon resistance thermometer (CRT) \cite{CRT}
to measure the change of liquid temperature.
CRT is calibrated at
the boiling ($T_{\mathrm{b}}$) and triple ($T_{\mathrm{t}}$) points
of liquid N$_{2}$. Between the two points, the
resistance-temperature relation is interpolated by a linear function, giving the maximum
calibration error of $\pm$\, 0.2 K. The thermometer is located away from the pipe end in order
to measure the mean temperature of liquid.

In order to confirm that the cooling of liquid is caused by the internal evaporation,
we monitored the evaporation rate by measuring the liquid weight with an electronic scale
as shown in the photograph (Fig. \ref{fig:2}(c)).
The precision of the electronic scale is $\pm$0.1 g.

\section{Results and Discussions}
\begin{figure}[thbp]
\begin{center}
\includegraphics[width=100mm]{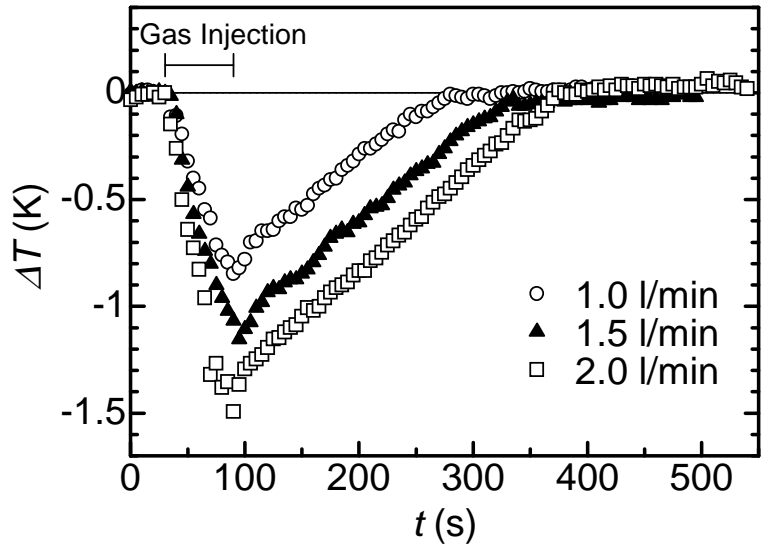}
\end{center}
\caption{The temperature changes of liquid N$_{2}$ when He gas is
injected at a constant flow rate denoted. The gas injection starts at
$t=$ 30 s and ends at $t=$ 90 s.}
\label{fig:3}
\end{figure}

Figure \ref{fig:3} shows the measured temperature
changes by the gas injection at flow rates of 1.0, 1.5 and 2.0 $\ell$/min for 60 s. The liquid temperature drops
down while He gas is blown. After stopping the injection, it rises back to $T_{\mathrm{b}}$ at a
constant speed. With increasing the flow rate, the temperature drop becomes larger,
resulting in longer boiling suppression. It was observed that the boiling restarts
just when the temperature returns
to $T_{\mathrm{b}}$ which is unchanged from that before the injection.

\begin{figure}[thbp]
\begin{center}
\includegraphics[width=100mm]{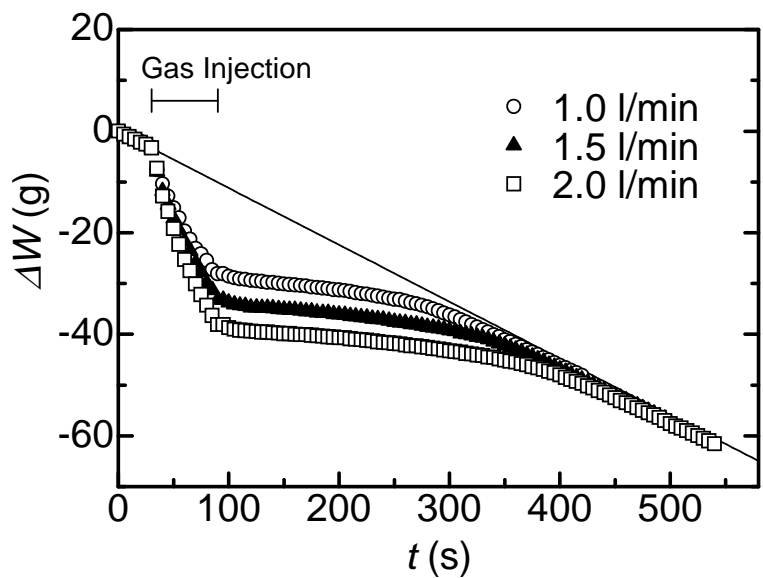}
\end{center}
\caption{The weight changes of liquid N$_{2}$ when He gas is
injected at a constant flow rate denoted.
The solid line is the weight change for the steady-state evaporation
under the ambient heat leak.}
\label{fig:4}
\end{figure}

The measured time evolutions of the liquid weight are shown in
Fig. \ref{fig:4}. 
The data shown in Figs. \ref{fig:3} and \ref{fig:4} were taken simultaneously.
Without the gas injection, liquid
N$_{2}$ evaporates steadily at a constant rate under a constant heat flow into the dewar
from the environment. The evaporation rate increases by a factor of 3-5 after the gas
injection starts at $t=30$ s. After the injection is stopped, the evaporation rate
becomes smaller than that in the steady state because the ambient heat leak is
absorbed by the cooled liquid N$_{2}$. It returns to the steady-state speed
and the boiling restarts when the temperature comes back to $T_{\mathrm{b}}$
at $t = 270$, 320 and 360 s for 1.0, 1.5 and 2.0 $\ell$/min, respectively.
This weight measurement demonstrates unambiguously that liquid
N$_{2}$ is cooled by the forced evaporation.

Let us calculate the temperature drop ($\Delta T$) due to this internal evaporation
using the data at the flow rate of 2.0 $\ell$/min in Fig. \ref{fig:4}.
At $t=90$ s, the excess weight loss of liquid N$_{2}$
beyond the natural evaporation is $\Delta W=3\times 10^{-2}$ kg.
Since the latent heat per unit mass of N$_{2}$ is $L=200$ kJ/kg,
the heat taken away by the forced evaporation is
$\Delta Q=L\Delta W=6$ kJ.
Specific heat of liquid N$_{2}$ ($c_{\mathrm{N}_{2}}$) and glass ($c_{\mathrm{G}}$)
is 2 and 0.7 kJ/kg$\cdot$K, and
the mass of liquid N$_{2}$ in the dewar ($M_{\mathrm{N}_{2}}$) and
the inner wall of glass dewar($M_{\mathrm{G}}$)
is 1.2 and 1 kg, respectively.
Thus, the total heat capacity is
$C=c_{\mathrm{N}_{2}}M_{\mathrm{N}_{2}}+c_{\mathrm{G}}M_{\mathrm{G}}=3.1$ kJ/K.
A part of $\Delta Q$ should be used to cool the injected He gas
from 300 K to approximately 77 K.
Such heat is calculated as
$\Delta Q_{\mathrm{He}}=c_{\mathrm{He}}M_{\mathrm{He}}\times(300-77)=0.42\;\mathrm{kJ}$,
where $c_{\mathrm{He}}(=5.2\;\mathrm{kJ}/\mathrm{kg}\cdot\mathrm{K})$ is $T$-independent
specific heat of He and $M_{\mathrm{He}}(=3.6\times10^{-4}\;\mathrm{kg})$
is the weight of the injected He gas.
Therefore, the temperature drop is calculated as
$\Delta T=(\Delta Q-\Delta Q_{\mathrm{He}})/C=1.8$ K.
This is roughly in accordance with the measured temperature drop (=1.4 K) at $t=90$ s,
which demonstrates the relevance of the forced evaporation mechanism quantitatively.

\begin{figure}[thbp]
\begin{center}
\includegraphics[width=100mm]{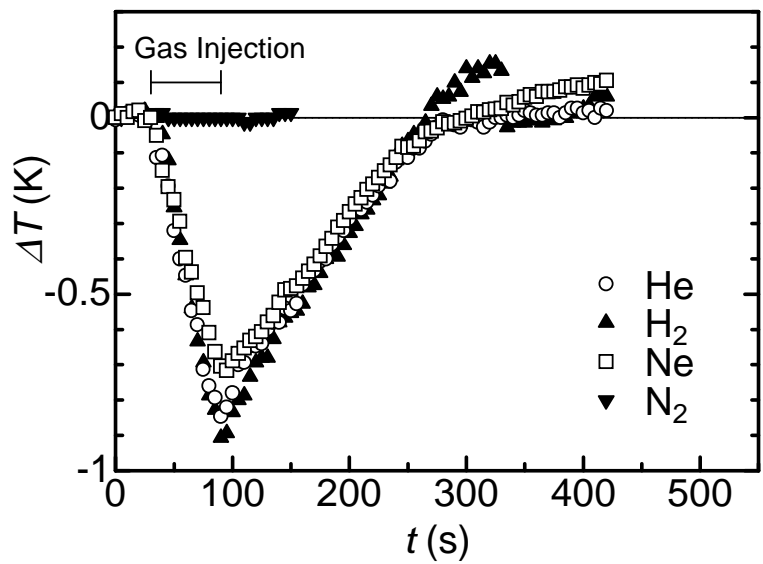}
\end{center}
\caption{The temperature changes of liquid N$_{2}$ when various kinds of gas are
injected at a constant flow rate of 1.0 $\ell$/min.}
\label{fig:5}
\end{figure}

Next we made temperature measurements with H$_{2}$, Ne and N$_{2}$ gases at a constant
flow rate of 1.0 $\ell$/min. It is difficult to conduct this experiment with other kinds of
gas with higher boiling points than N$_{2}$, since they are liquefied
into liquid N$_{2}$. The results are shown in Fig. \ref{fig:5}. The cooling rates during the
gas injection of Ne and H$_{2}$ are almost the same as that of He. On the other hand,
the temperature drop is not observed with N$_{2}$ injection.
This shows that the role of injected gas is to increase the liquid
surface area and to displace the
evaporated gas to the atmosphere efficiently regardless of kind
of injected gas.

For H$_{2}$ gas in Fig. \ref{fig:5},
the temperature rises above $T_{\mathrm{b}}$ and suddenly returns to $T_{\mathrm{b}}$ at
$t = 340$ s, at which boiling restarts. This is overheating, which is sometimes
observed in the case of less nucleation centers,
i.e., when liquid N$_{2}$ is not contaminated by ice grains condensed from
the atmosphere and the inner surface of the dewar is clean enough.

\begin{figure}[thbp]
\begin{center}
\includegraphics[width=100mm]{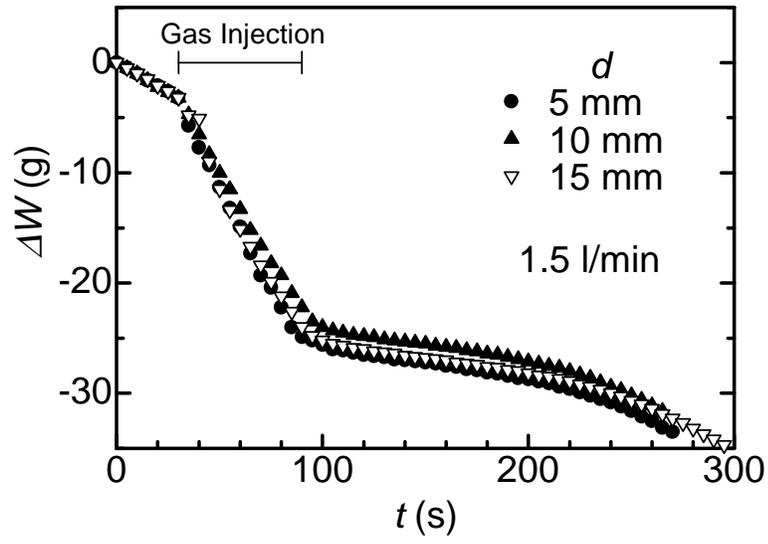}
\end{center}
\caption{The weight changes of liquid N$_{2}$ when He gas is
injected through pipes of different diameters ($d$).
The flow rate is fixed at 1.5 $\ell$/min.}
\label{fig:6}
\end{figure}

We found that the cooling power does not depend on
the size of bubbles by changing the pipe diameter ($d=5$, 10, 15 mm) with a fixed flow rate
as shown in Fig. \ref{fig:6}.
This indicates that N$_{2}$ vapor pressure in the bubbles is saturated
before arriving at the liquid surface.
The N$_{2}$ boiling suppression was also observed
by replacing the air above the liquid N$_{2}$ surface with He gas
as was reported in Ref. \cite{Minkoff},
although the temperature drop is smaller than that caused by the injection.

\begin{figure}[thbp]
\begin{center}
\includegraphics[width=100mm]{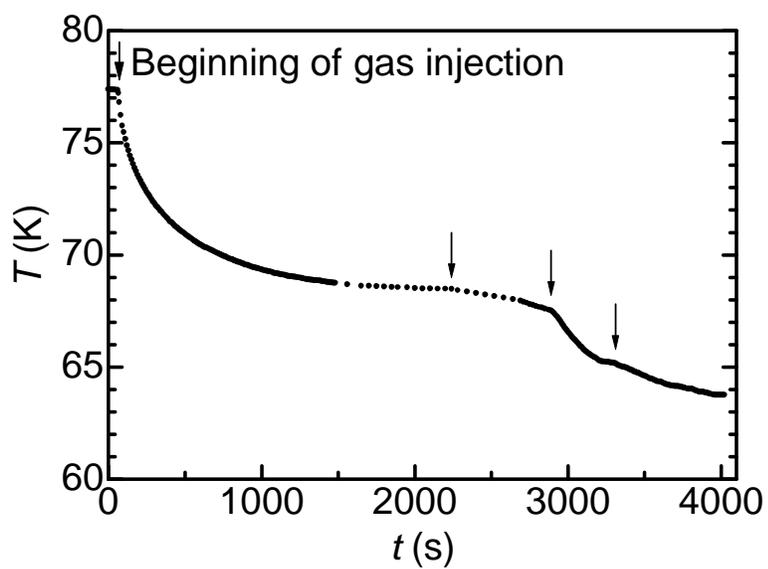}
\end{center}
\caption{The attempt to cool liquid N$_{2}$ as much as possible,
using a metal dewar.
The flow rate is increased stepwise at the times indicated by the arrows.
The highest flow rate is 5.0 $\ell$/min.}
\label{fig:7}
\end{figure}

Finally, we tried to lower the temperature of liquid N$_{2}$
by the forced evaporation method as much as possible.
As shown in Fig. \ref{fig:1}, the vapor pressure of N$_{2}$
at the triple point
is as high as $1.25\times 10^{4}$ Pa, and the temperature difference
between $T_{\mathrm{t}}=63.2$ K and $T_{\mathrm{b}}=77.4$ K
is only 14.2 K. Hence, in principle, it is possible to solidify N$_{2}$ by this method,
which was demonstrated by the previous workers \cite{Xu}. We used a metallic container
rather than the transparent dewar in order to decrease the ambient heat leak,
and He gas was precooled before injection.
Figure \ref{fig:7} is the result of this cooling experiment.
Here, the flow rate is increased stepwise at $t=2240,2890$ and 3310 s
indicated by the arrows in Fig. \ref{fig:7}
up to 5.0 $\ell$/min at the end.
We reached the lowest temperature of $63.7\pm 0.2$ K,
which is very close to $T_{\mathrm{t}}$
as shown in Fig. \ref{fig:7}. The boiling was suppressed
for more than 150 min after stopping the injection.
Considering the fact that CRT was placed near the pipe end and
the temperature drop was saturated,
it is plausible that we reached the triple point actually.
However, solid N$_{2}$ was not identified at least by eyes.
This is presumably because the liquid was heavily contaminated with the ice dust.

\section{Conclusion}
We confirmed that the N$_{2}$ boiling suppression with He gas injection
is caused by the liquid temperature drop due to the forced evaporation.
This was made by measuring the liquid temperature and weight simultaneously.
This phenomenon is practically useful for experiments where
one has to minimize vibrational and optical disturbances.

\section*{Acknowledgements}
The authors are thankful to Y. Ando, E. Iyoda, Y. Ota, K. Oda, M. Kuroda, C. Matsui
and T. Misumi for their contributions to the experiment and valuable discussions.

\newpage
\noindent

\newpage
\noindent

\newpage
\noindent

\newpage
\noindent

\newpage
\noindent

\newpage
\noindent

\newpage
\noindent
\end{document}